\documentstyle[aps,epsfig]{revtex}

\begin{document}

\draft 

\preprint{to appear in Phys. Rev. Lett.}

\title{Kaon production in heavy-ion collisions
and maximum mass of neutron stars}

\bigskip
\author{G.Q. Li, C.-H. Lee, and G.E. Brown}
\address{Department of Physics, State University of New
York at Stony Brook, Stony Brook, New York 11794}
\maketitle
  
\begin{abstract}
We determine an `empirical' kaon dispersion relation
by analysing and fitting recent experimental data on kaon production
in heavy-ion collisions. We then investigate its effects on 
hadronic equation of state at high densities and on
neutron star properties. We find that the maximum mass of neutron stars
can be lowered by about 0.4$M_\odot$, once kaon condensation as 
constrained by our empirical dispersion relation is introduced. 
We emphasize the growing interplay between hadron physics, 
relativistic heavy-ion physics and the physics of compact objects 
in astrophysics.
\end{abstract}


There is currently growing interplay between the physics of
hadrons (especially the properties of hadrons in 
dense matter which might reflect spontaneous
chiral symmetry breaking and its restoration), the physics
of relativistic heavy-ion collisions (from which one might 
extract hadron properties in dense matter), and the
physics of compact objects in astrophysics (which needs
as inputs the information gained from the first two fields).
A notable example is the kaon ($K$ and ${\bar K}$), which, 
being a Goldstone boson with strangeness, 
plays a special role in all of the three fields mentioned.

Ever since the pioneering work of Kaplan and Nelson \cite{kap86}
on the possibility of kaon condensation in nuclear matter,
much theoretical effort has been devoted to
the study of kaon properties in dense matter.
Brown {\it et al.} \cite{lee96}
have carried out a detailed study of free-space and in-medium
kaon-nucleon scattering using chiral perturbation
theory. Yuba {\it et al.} studied kaon in-medium
properties based on phenomenological off-shell meson-nucleon
interactions \cite{kubo93}. Weise and collaborators
investigated this problem using the Nambu$-$Jona-Lasinio
model, treating the kaon as a quark-antiquark excitation
\cite{weise94}. Recently they have extended the chiral perturbation
calculation to include the coupled-channel effects which are
important for the $K^-$ meson \cite{weise96}.
Another type of study, which is based on the extension
of the Walecka mean-field model from SU(2) to SU(3),
was pursued by Schaffner {\it et al.} \cite{sch94}
and Knorren {\it et al.} \cite{kno95}.
Although quantitatively results from these different models
are not identical, qualitatively, a  
consistent picture has emerged; namely, in nuclear matter
the $K^+$ feels a weak repulsive potential, whereas
the $K^-$ feels a strong attractive potential. 

Measurements of kaon spectra and flow have been 
carried out in heavy-ion collisions at SIS (1-2 AGeV), 
AGS (10 AGeV),  and SPS (200 AGeV) energies \cite{qm96}. 
By comparing transport model predictions with experimental 
data, one can learn not only the global reaction dynamics,
but more importantly, the kaon properties in dense
matter. Of special interest is kaon production in
heavy-ion collisions at SIS energies, as it has been shown
that particle production at subthreshold energies 
is sensitive to its properties in dense matter
\cite{sub96}. Recently, high quality data concerning 
$K^+$ and $K^-$ production in heavy-ion collisions at 
SIS energies have been published by
the KaoS collaboration at GSI \cite{kaos}. 
The KaoS data show that the $K^-$ yield at 1.8 AGeV 
(projectile nucleus kinetic energy in the laboratory frame)
agrees roughly
with the $K^+$ yield at 1.0 AGeV. This is a nontrivial observation.
These beam energies were purposely chosen, such that 
the Q-values for $NN\rightarrow NK\Lambda$ and 
$NN\rightarrow NNK{\bar K}$ are both  
about -230 MeV.  Near their respective production thresholds,
the cross section for the $K^-$ production in proton-proton 
interactions is one to two orders of magnitude smaller than 
that for $K^+$ production \cite{data}. In addition, antikaons
are strongly absorbed in heavy-ion collisions, which should further
reduce the $K^-$ yield. The KaoS results of $K^-/K^+\sim 1$
indicate thus the importance of kaon medium effects which 
act oppositely on $K^+$ and $K^-$ production in nuclear
medium.
 
Studies of neutron star properties also have a long history. A recent 
compilation by Thorsett quoted by Brown \cite{brown}
shows that well-measured neutron star masses are all less
than 1.5$M_\odot$. On the other hand, most of the theoretical 
calculations based on conventional nuclear equations of state 
(EOS) predict a maximum neutron state mass above 2$M_\odot$. 
The EOS can, therefore, be substantially softened without running
into contradiction with observation. 
Various scenarios have been proposed that can lead to a soft
EOS, including the high-order self-interactions of the vector
field \cite{mull96}, the possibility of kaon condensation \cite{brown94},
the existence of hyperons \cite{kno95,glen92,sch96},
and the transition to quark matter \cite{glen95}. 
All these possibilities need to be examined against the available
empirical information from, e.g., relativistic heavy-ion
collisions.

The chief aim of this paper is to determine, from the recent
KaoS data on kaon production, together with previous analysis
of nucleon flow, kaon flow and dilepton
spectra \cite{eos,fopi95,ceres,gqli96}, an `empirical' kaon
dispersion relation in dense matter.
We will show that these data are consistent with 
the scenario that the $K^+$ feels a weak repulsive potential
and the $K^-$ a strong attractive potential, as predicted
by the chiral perturbation calculation.
We then study the effects of this empirical dispersion relation
on the possibility of kaon condensation and on the
neutron star properties. We find that 
$K^-$ condensation happens at about 3$\rho_0$, and the 
maximum mass of neutron stars is lowered by about 0.4$M_\odot$
once the kaon condensation is introduced. 
These values change by about 20\% when different nuclear
equations of state are used (see Ref. \cite{thor94} for
a detailed discussion).

We use the relativistic transport model for the description of
heavy-ion collisions and for the calculation of kaon production
\cite{sub96}. The nucleon dynamics is governed by the chiral Lagrangian
recently developed by Furnstahl, Tang and Serot \cite{fst},
which is derived using dimensional analysis, naturalness 
arguments, and provides a very good description of nuclear 
matter and finite nuclei. Furthermore, a recent analysis
\cite{gqli96} showed that this model reproduces nicely
the nucleon flow \cite{eos} and dilepton spectra \cite{ceres}
in heavy-ion collisions, indicating that its extrapolation
to 2-3$\rho_0$ is consistent with empirical information.
We will thus use this model as our basis for the 
determination of kaon dispersion relation and for the
analysis of neutron star properties.  

In the mean-field approximation, the energy density 
for the general case of asymmetric nuclear matter 
is given by
\begin{eqnarray}
\varepsilon _N& = & {2\over (2\pi )^3} \int _0^{K_{fp}} 
d{\bf k} \sqrt {{\bf k}^2+m_N^{*2}} +{2\over (2\pi )^3} 
\int _0^{K_{fn}} d{\bf k} \sqrt {{\bf k}^2+m_N^{*2}} \nonumber\\
 & + & W\rho +R {1\over 2}(\rho_p-\rho_n) -{1\over 2C_V^2}W^2
- {1\over 2C_\rho^2}R^2 + {1\over 2C_S^2}\Phi^2 \nonumber \\
 & +& {S^{\prime 2}\over 4C_S^2}d^2\left\{\left(1-{\Phi \over S^\prime}
\right)^{4/d}\left[{1\over d}{\rm ln}\left(1-{\Phi \over S^\prime}
\right) - {1\over 4}\right]+{1\over 4}\right\}
-{\xi\over 24}W^4 - {\eta \over 2C_V^2}{\Phi \over S^\prime}W^2. 
\end{eqnarray}
The nucleon effective mass $m_N^*$ is related to its scalar
field $\Phi$ by $m_N^*=m_N-\Phi$. $W$ and $R$ 
are the isospin-even and isospin-odd vector potentials,
respectively. The last three terms give the self-interactions of
the scalar field, the vector field, and the coupling between
them. The meaning and values of various parameters in Eq. (1)
can be found in \cite{fst}. In this work, we use the
parameter set T1 listed in Table 1 of \cite{fst}.

\begin{figure}
\begin{center}
\centerline{\epsfig{file=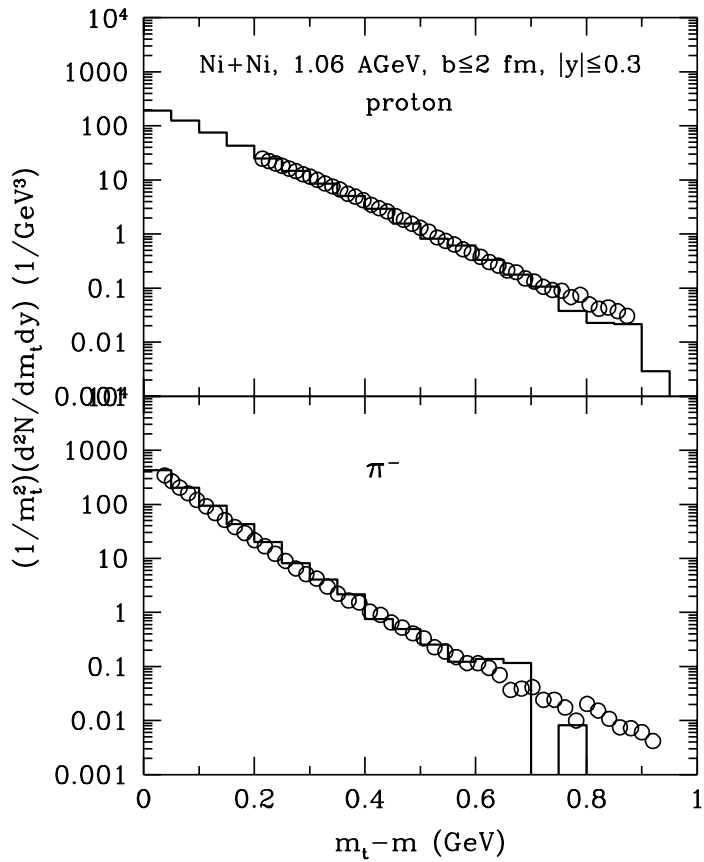,height=5in,width=5in}}
\caption{Proton and $\pi^-$ transverse mass spectra in central
Ni+Ni collisions at 1.06 AGeV.}
\end{center}
\end{figure}

From Eq. (1), we can derive a relativistic transport model for heavy-ion
collisions. At SIS energies, the colliding system consists
mainly of nucleons, delta resonances, and pions. While medium
effects on pions are neglected,
nucleons and delta resonances propagate in a common
mean-field potential according to the Hamilton equation of
motion,
\begin{eqnarray}
{d{\bf x}\over dt} = {{\bf p}^*\over E^*}, \;\;\;
{d{\bf p}\over dt} = - \nabla _x (E^*+W),
\end{eqnarray}
where $E^*=\sqrt {{\bf p}^{*2} + m^{*2}}$.
These particles also undergo stochastic two-body
collisions \cite{sub96}. In Fig. 1 we compare our results
for proton and pion transverse mass spectra in central
Ni+Ni collisions with experimental data from the FOPI 
collaboration \cite{fopi}. The nice agreement with the
data provides further support to the use of the chiral 
Lagrangian of \cite{fst} in the present analysis.

In heavy-ion collisions at SIS energies, kaons can 
be produced from pion-baryon and baryon-baryon collisions.
For the former we use cross sections obtained in the resonance model
by Tsushima {\it et al.} \cite{fae94}. For the latter the
cross sections obtained in the one-boson-exchange model
of Ref. \cite{likoc97} are used. Both models describe the
available experimental data very well. For antikaon production
from pion-baryon collisions we use the parameterization
proposed by Sibirtsev {\it et al.} \cite{sib97}. For
baryon-baryon collisions, we use a somewhat different 
parameterization, which describes the experimental data better,
than Ref. \cite{sib97}. In addition, the antikaon
can also be produced from strangeness-exchange processes
such as $\pi Y\rightarrow {\bar K}N$ where $Y$ is a hyperon. 
The cross sections for these processes are obtained from the 
inverse ones, ${\bar K}N\rightarrow \pi Y$, by the 
detailed-balance relation. The latter cross sections,
together with the ${\bar K}N$ elastic and absorption cross
sections, are parameterized based on the available experimental
data \cite{data}. The details about elementary cross sections,
the transport model, and the neutron star calculation 
will be given elsewhere \cite{lilee97}.
  
From the chiral Lagrangian the kaon and antikaon in-medium energies
can be written as \cite{lee96}
\begin{eqnarray}
\omega _K=\left[m_K^2+{\bf k}^2-a_K\rho_S
+(b_K \rho )^2\right]^{1/2} + b_K \rho 
\end{eqnarray}
\begin{eqnarray}
\omega _{\bar K}=\left[m_K^2+{\bf k}^2-a_{\bar K}\rho_S
+(b_K \rho )^2\right]^{1/2} - b_K \rho 
\end{eqnarray}
where $b_K=3/(8f_\pi^2)\approx 0.333$ GeVfm$^3$, 
$a_K$ and $a_{\bar K}$  are two parameters
that determine the strength of the attractive scalar potential for
kaon and antikaon, respectively. If one considers only the
Kaplan-Nelson term, then $a_K=a_{\bar K}=\Sigma _{KN}/f_\pi ^2$.
In the same order, there is also the range term which acts differently
on kaon and antikaon, and leads to different scalar attractions.
Since the exact value of $\Sigma _{KN}$ and the size of
the higher-order corrections are still under intensive
debate, we take the point of view that $a_{K,{\bar K}}$ 
can be treated as free parameters and try to constrain 
them from the experimental observables. Since the $KN$ interaction is
relatively weak, impulse approximation should be reasonable at low
densities. This provides some constraints on $a_K$. We find that
$a_K\approx 0.22$ GeV$^2$fm$^3$, corresponding to 
$\Sigma _{KN} \approx 400$ MeV, gives a repulsive $K^+$ potential 
of about 20 MeV at normal nuclear matter density, 
slightly smaller than the 25 MeV found in \cite{koch95}.
We will show later that this value also gives a good fit to 
the $K^+$ spectra in heavy-ion collisions.

From the chiral Lagrangian we can also derive equations of motion for
kaons \cite{liko95}
\begin{eqnarray}
{d{\bf x}\over dt}= {{\bf p}^*\over \omega _{K,{\bar K}}\mp b_k\rho },
\;\;\; {d{\bf p} \over dt}= - \nabla _x \omega _{K,{\bar K}},
\end{eqnarray}
The minus and the plus sign correspond to kaon and antikaon, 
respectively. 

For $K^+$ and $K^-$ production in heavy-ion collisions,
we consider two scenarios; namely, with and without
kaon medium effects. As mentioned, we use $a_K\approx 0.22$
GeV$^2$fm$^3$ for $K^+$. For $K^-$, we adjust $a_{\bar K}$ such that we 
achieve a good fit to the experimental $K^-$ spectra.
We find $a_{\bar K}\approx 0.45$ GeV$^2$fm$^3$, which leads to
a $K^-$ potential of about -110 MeV at normal nuclear 
matter density. This is somewhat smaller, in magnitude, than the 
`best' value of $-200 \pm 20$ MeV extracted from kaonic 
atoms \cite{gal94}. The latter value, however,
depends sensitively on the extrapolation procedure from
the surface of nuclei to their interiors \cite{gal94}. On
the other hand, kaon production in heavy-ion collisions and
neutron star calculations are sensitive chiefly to higher densities
(2-3$\rho_0$), where our value should be more relevant.
 
\begin{figure}
\begin{center}
\epsfig{file=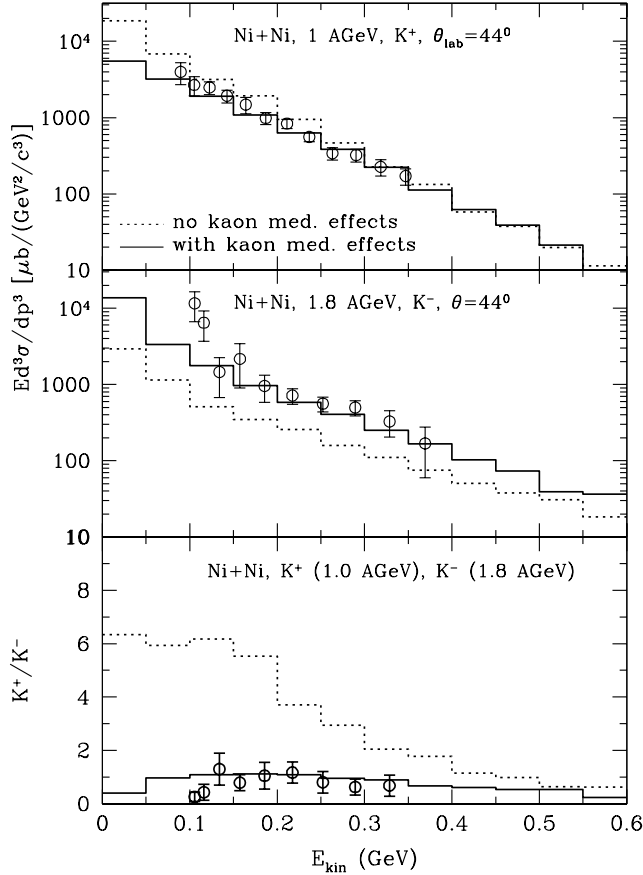,height=5in,width=5in}
\caption{$K^+$ (upper window), $K^-$ (middle window), and $K^+/K^-$
(lower window) kinetic energy spectra in Ni+Ni collisions.}
\end{center}
\end{figure}

The results for the $K^+$ and $K^-$ kinetic energy 
spectra are shown Fig. 2. The solid and dotted histograms give
the results with and without kaon medium effects, respectively.
The open circles are the experimental data from the KaoS collaboration
{\cite{kaos}}. For the $K^+$, it is seen that the results with 
kaon medium effects are in good agreement with the data, while
those without kaon medium effects slightly overestimate
the data. We note that the kaon feels a slightly repulsive potential;
thus the inclusion of the kaon medium effects reduces the kaon yield.
For the $K^-$, it is seen that without medium effects, our
results are about a factor 3-4 below the experimental data.
With the inclusion of the medium effects which reduces the
antikaon production threshold, the $K^-$ yield increases by 
about a factor of 3 and our results are in good agreement 
with the data. This is similar to the findings of Cassing 
{\it et al.} \cite{cassing}. For both $K^+$ and $K^-$, the
differences between the two scenarios are most pronounced at low
kinetic energies. The experimental data at these momenta
will be very useful in discriminating the two scenarios.

The effects of kaon and antikaon mean-field potentials 
can be more clearly seen by looking at their ratio as a 
function of the kinetic energy, which is shown in the lower 
window of Fig. 2. Without kaon medium effects, the $K^+/K^-$
ratio decreases from about 7 at low kinetic energies to
about 1 at high kinetic energies, which is in complete disagreement
with the data. Since the antikaon absorption cross section by nucleons
becomes large at low momentum, low-momentum
antikaons are more strongly absorbed than 
high-momentum ones. This makes the $K^+/K^-$ ratio increase
with decreasing kinetic energies. When medium effects are
included, we find that the $K^+/K^-$ ratio is almost
unity in the entire kinetic energy region,
which agrees very well with the data. 
The shapes of the $K^+$ and $K^-$ spectra change in 
opposite ways in the presence of their mean-field potential.
Kaons are `pushed' to high momenta by the repulsive potential,
while antikaons are `pulled' to low momenta. 
The good description of the $K^+/K^-$
ratio, together with the fit to the kaon flow in heavy-ion
collisions \cite{fopi95,liko95}, gives us confidence that our 
in-medium kaon and antikaon dispersion relations are reasonable.

\begin{figure}
\begin{center}
\epsfig{file=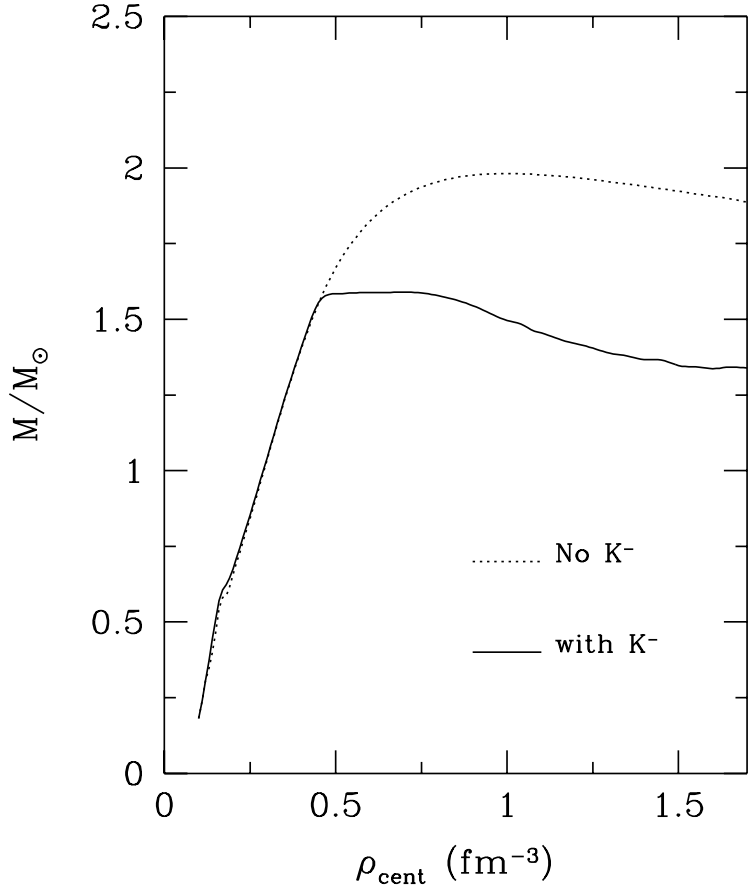,height=5in,width=5in}
\caption{Neutron star mass as a function of central density.
The solid and dotted lines are obtained with and
without $K^-$ condensation.}
\end{center}
\end{figure}

We take the antikaon dispersion relation constrained by the
heavy-ion data as empirical indication of an attractive
antikaon potential in dense matter. We combine this
with the energy density of Eq. (1) for nuclear matter,
and calculate neutron star properties. We find that
around $3\rho_0$, the effective mass of the $K^-$
drops below the chemical potential of the electron, indicating
the onset of kaon condensation. After this the $K^-$ density
increases rapidly, leading to a large proton fraction in
the neutron star. The results for neutron star mass as function
of central density is shown in Fig. 1, where the solid and
dotted curves give the results with and without $K^-$,
respectively. It is seen that the maximum mass of the neutron stars is
reduced by about 0.4$M_\odot$ with the introduction of
the kaon condensation. Both the critical density
for kaon condensation and the amount of lowering in the maximum 
neutron star mass change by about 20\% when different nuclear equations
of state are used. The exact value of the maximum neutron star
mass, on the other hand, depends more sensitively on the particular
nuclear EOS used, as discussed in \cite{mull96}.

It should be mentioned that alternative explanations for lowering
the maximum neutron star mass have been proposed. Of particular
interest is the introduction of $\Sigma^-$ hyperons in \cite{glen92}.
This is a complementary rather than competing scenario,
since the $\Sigma^-$-particle$-$neutron-hole state 
has a $p$-wave coupling to the $K^-$. A unification of the 
scenarios can be achieved by introducing the `kaesobar', a linear 
combination of $K^-$ and $\Sigma^-$-particle$-$neutron-hole  
\cite{brown98}, but the results of the present work will not
be strongly modified.

In summary, we studied $K^+$ and $K^-$ production 
in Ni+Ni collisions at 1-2 AGeV, based on the relativistic 
transport model including the strangeness degrees of freedom. 
We found that the recent experimental data from the KaoS
collaboration are consistent with the predictions of chiral
perturbation theory that the $K^+$ feels a weak repulsive potential
and the $K^-$ feels a strong attractive potential in the nuclear 
medium. Using the kaon in-medium properties constrained
by heavy-ion data, we have studied the possibility of kaon condensation
and its effects on neutron star properties.
The critical density for kaon condensation was found to be
about 3$\rho_0$, and the maximum mass of neutron
stars was found to be redduced by about 0.4$M_\odot$
once kaon condensation is introduced. 
We have emphasized the growing interdependence
of hadron physics, relativistic heavy-ion physics
and the physics of compact stars in astrophysics.

We are grateful to C.M. Ko, T.T.S. Kuo, M. Prakash, and M.
Rho for useful discussions. We also thank N. Herrmann 
and P. Senger for useful communications.
This work is supported in part by the 
Department of Energy under Grant No. DE-FG02-88ER40388.

\end{document}